\documentclass[paper,12pt]{article}
\usepackage[centertags]{amsmath}
\usepackage{amsfonts} \usepackage{amssymb} \usepackage{amsthm}
\usepackage{graphicx}
\usepackage{bbm}
\usepackage{pstricks}
\let\a=\alpha

         \let\w=\omega

       \let\C=\Chi    

\def\cI{{\cal I}}

\def\cN{{\cal N}}

\def\cS{{\cal S}}

\newcommand{\be}{\begin{equation}}
\newcommand{\ee}{\end{equation}}
\newcommand{\bea}{\begin{eqnarray}}
\newcommand{\eea}{\end{eqnarray}}
\newcommand{\ba}{\begin{array}}
\newcommand{\ea}{\end{array}}
\def\nn{\nonumber}

\newcommand{\ft}[2]{{\textstyle\frac{#1}{#2}}}
\newcommand{\Z}{{\mathbb Z}}

\newcommand{\C}{{\mathbb C}}

\newcommand{\one}{\mathbbm{1}}
\begin{document}
\begin{titlepage}
\begin{flushright}
{ROM2F/2009/06}\\
\end{flushright}
\begin{center}
{\large \sc Dynamical supersymmetry breaking \\ from unoriented
D-brane instantons }\\
\vspace{1.0cm}
{\bf Massimo Bianchi}, {\bf Francesco Fucito} and {\bf Jose F. Morales}\\
{\sl Dipartimento di Fisica, Universit\'a di Roma ``Tor Vergata''\\
 I.N.F.N. Sezione di Roma II\\
Via della Ricerca Scientifica, 00133 Roma, Italy}\\
\end{center}
\vskip 2.0cm
\begin{center}
{\large \bf Abstract}
\end{center}
We study the non-perturbative dynamics of an unoriented
$\Z_5$-quiver theory of GUT kind  with gauge group $U(5)$ and
chiral matter. At strong coupling the  non-perturbative dynamics
is described in terms of  set of baryon/meson variables satisfying
a quantum deformed constraint. We compute the effective
superpotential of the theory and show that it admits a line of
supersymmetric vacua and a phase where supersymmetry is
dynamically broken via gaugino condensation. \vfill
\end{titlepage}

\tableofcontents

\setcounter{section}{0}

\section{Introduction}

 The search for mechanisms of supersymmetry breaking in string and
field theories has a long history. Out of the many alternatives,
the idea that supersymmetry be broken dynamically remains one of
the most attractive choices.  In this framework, supersymmetry
remains a symmetry  of the effective action even after
non-perturbative effects are included and is only spontaneously
 broken by the vacuum choice
 \cite{Witten:1981nf,Witten:1982df,Amati:1988ft,Affleck:1984mf,Affleck:1984xz}.
This is in contrast with scenarios based on explicit supersymmetry breaking 
where the loss of supersymmetry has more dramatic
effects and can be hardly followed in a controllable way.  Still
the conditions under which  supersymmetry is broken
  are very restrictive and  the search for string realizations of such scenarios
  remains a far from obvious task  (see \cite{Berenstein:2005xa}
\nocite{Franco:2005zu,Bertolini:2005di,Diaconescu:2005pc,Florea:2006si,Aharony:2007db,Buican:2008qe,Cvetic:2008mh,Heckman:2008es,Dudas:2008qf}
-\cite{Marsano:2008jq} for recent
  contributions in the subject).

  In this paper we study in details an example of
 a ${\cal N}=1$ chiral model, built out of D3-branes at a
 $\C^3/\Z_5$ singularity, exhibiting  dynamical supersymmetry breaking.
At low energy the D-brane dynamics is described by a quiver gauge
theory with gauge groups and matter representations
specified by the choice of fractional D3-branes.
We consider  a $\Z_5$ quiver  leading to a gauge theory with gauge group
 $U(5)\times U(1)$, two flavours of chiral matter in the ${\bf 10}$ dimensional 
representation of the gauge group, three flavours in the ${\bf \bar 5}$, one flavour
in the  ${\bf 5}$ and two flavour singlets.
We present a complete description of the
non-perturbative dynamics of the gauge theory. The strong coupling
dynamics of the theory is described in terms of a set of gauge
invariant composites, `baryons' and `mesons', satisfying a quantum
deformed constraint.  The vacuum structure of the theory depends
very much on whether Fayet-Iliopoulos (FI) terms are turned on or
not. In the absence of FI terms a supersymmetric line of vacua is
found. Turning on FI terms, the theory undergoes a Higgs
mechanism, whereby supersymmetry is dynamically broken  by
instanton effects.

   Our analysis will be local and will not address the issue of how
   to embed the quiver gauge
theory in a global vacuum configuration of (unoriented) D-branes.
 At low energies the physics near the singularity is efficiently
captured by the local description of the singularity and
constraints such as global tadpole cancellation can  be neglected
at a first look. We assume that a global description exist and
that the other (unoriented) D-branes needed for global R-R tadpole
cancellation are located far from the singularity.
   We remark that although $T^6$ tori do not admit
$\Z_5$ actions compatible with residual supersymmetry,
`supersymmetric' $\C^3/\Z_5$ singularites may show up in Calabi
Yau compactifications as for instance in the $\Z_5$ quotient of
the quintic\footnote{We thank E.~Witten for suggesting us this
embedding.}.

The paper is organized as follows. In section 2, we briefly review
the construction of unoriented $\Z_n$ quiver gauge theories and
identify a  grand unification (GUT) like model with gauge group $U(5)\times U(1)$ and
chiral matter content. In section 3 we describe the strong
coupling dynamics of the theory.    In section 4 we test our
strong coupling description by matching the anomalies of the
effective theory described in terms of baryon/meson composites
with those computed in terms of elementary
 fields. In section 5 we identify the relevant instanton moduli
 space and interactions and
 sketch the D-instanton
 derivation of the non-perturbative superpotential.
 In section 6 we summarize our results.

\section{Unoriented $\Z_n$ quivers}

The low-energy dynamics of $N$ D3-branes in flat spacetime is
described by a $\cN =4$ supersymmetric gauge theory in four
dimensions with gauge group  $U(N)$.  In the $\cN=1$ language this
theory comprises a vector superfield $V$ and  three chiral fields
$\Phi^I$, $I=1,2,3$ transforming in the adjoint representation of
$U(N)$. In addition one has a cubic superpotential of the form 
\be
W_{\rm tree}=  {\rm Tr} \, \Phi^1[\Phi^2,\Phi^3]
\label{wtree} 
\ee 
If one puts D3-branes at a singularity
$\C^3/\Gamma$, with $\Gamma$ a discrete subgroup of $SU(3)$,  the
low-energy dynamics is described by an $\cN =1$ quiver gauge
theory with bi-fundamental matter specified by the intersection
matrix of cycles on $\C^3/\Gamma$ and a cubic superpotential
descending from (\ref{wtree}).

We take  $\Gamma = \Z_n$ acting on the $\C^3$ coordinates
$z_I$ as \be \Z_n : \qquad     z^I \to \w^{q_I}\, z^I\qquad \omega
= e^{2\pi i/n} \qquad \sum_{I=1}^3 q_I=0~{\rm mod}~n
  \ee
   The orbifold group action
   on Chan-Paton indices can be represented by the  $\gamma_{g}$
   block diagonal matrix
\be
\gamma_{g} =\left(
\begin{array}{cccc}
    \one_{N_0\times N_0}    &  0 &... &... \\
 0 &  \w \,  \one_{N_1\times N_1}  &0 &  ... \\
  ...& 0 & ...&   0  \\
  ... &  ... & 0 & \w^{n-1} \, \one_{N_{n-1}\times N_{n-1}}   \\
\end{array}
\right) \ee with $N=\sum_a N_a$  the original number of branes and
$N_a$ the number of fractional D3-branes of type ``$a$''.
 The quiver gauge theory at the singularity is defined by restricting the ${\cal N}=4$  fields
  $V$, $\Phi^I$ to  $N\times N$ matrices satisfying the
  $\Z_n$-invariant conditions
 \be
\Z_n: \qquad V= \gamma_{g} \,V\, \gamma^\dagger_{g} \qquad
 \Phi^I = \w^{q_I}\, \gamma_{g} \,\Phi^I \, \gamma^\dagger_{g}
 \label{groupaction}
 \ee
 This results into block matrices with non-trivial components
  \bea
  V:  &&  \sum_{a=0}^{n-1} \, \quad N_a \times \bar N_{a}\nn\\
  \Phi^I : &&    \sum_{a=0}^{n-1} \, \quad N_a \times \bar N_{a+q_I}
  \label{cmatter}
\eea
 The resulting quiver gauge theory has then gauge group
 $ G = \prod_{a=0}^{n-1} U(N_a)$ and bi-fundamental matter.

In general, the matter content (\ref{cmatter}) is either non-chiral or
anomalous and therefore of very limited interest from the
phenomenological point of view.
 More appealing gauge theories can be realized by introducing
 $\Omega$-planes or equivalently
by quotienting the quiver by an anti-holomorphic involution
$\Omega$ that includes world-sheet parity.  $\Omega$ identifies
ingoing $\{ N_a \}$ and outgoing $\{ \bar N_a \}$ boundaries and
invariant fields  are defined by the restriction \be \Omega:
\qquad V= \epsilon\, \gamma_{\Omega} \,V\, \gamma^\dagger_{\Omega}
\qquad
 \Phi^I = -\epsilon\, \gamma_{\Omega} \,\Phi^I \, \gamma^\dagger_{\Omega}
 \label{groupaction2}
 \ee
  with $\epsilon$ a sign and $\gamma_{\Omega}$
  specifying the identification of boundaries.
   A canonical choice is given by
   taking\footnote{For $n$ even an extra choice is possible
 $
  \bar N_a=N_{n-a-1}
 $
 and $\gamma_{\Omega}$ a block matrix with non-trivial blocks  $\gamma_{\Omega} =\oplus_{a=0}^{n-1}  \one_{N_a\times N_{n-a-1}}$.}
\be
 \bar N_a=N_{n-a}
\ee
and
\be
\Omega: \qquad
 \gamma_{\Omega} =\small{ \left(
\begin{array}{cccccc}
    \one_{N_0\times N_0}    &  0&...& ... &... &0 \\
 0 & 0& ...&...& 0&  \one_{N_1\times N_1}   \\
  ...& ... & ...&...& \one_{N_2\times N_2} &   0  \\
  ...& ...&  ... &...&...&   ...  \\
   ...& 0&   \one_{N_{2}\times N_{2}} & 0 &...& ...   \\
 0&   \one_{N_{1}\times N_{1}} & 0& ... & ...&...   \\
\end{array}
\right)} \ee Moreover depending on the choice of R-R charge
$\epsilon=\pm$  of the $\Omega^{\pm}$-plane one has to
(anti)symmetrize the open strings with ends on image
stacks\cite{Bianchi:1991eu, Bianchi:1997rf, Witten:1997bs}. This
results into a quiver gauge theory with gauge group a product of
unitary (for $N_a$ complex) and orthogonal (symplectic) gauge
group components for $N_a$ real and $\epsilon=-(+)$.  In addition
chiral matter appear in bi-fundamental, symmetric or antisymmetric
representations of the gauge group according to
(\ref{groupaction2}).  Finally (local) twisted tadpole
cancellation, or equivalently cancellation of gauge anomalies,
constrains the choices of $N_a$ \cite{Bianchi:2000de,Aldazabal:2000sa}. More precisely, denoting by \be n_{
f,a\pm}=n_{f,a}\pm \bar n_{f,a} \quad\quad n_{S,a\pm}=n_{S,a}\pm
\bar n_{S,a}  \quad\quad n_{A,a\pm}=n_{A,a}\pm\bar n_{A,a} \ee the
chiral-antichiral (sum) difference of multiplets transforming in
the fundamental, symmetric or antisymmetric representation of a
$U(N_a)$ gauge group, the cancellation of gauge anomalies
requires\footnote{In our conventions: ${\rm tr}_{\bf R} T^a \{
T^b, T^c\}=A({\bf R}) d_{abc}$ and $A({\bf F})=\ft12$.  }
 \bea
 I_{a}&=& \sum_{\bf R_a}  A({\bf R_a})\nn\\
 &=& \frac{1}{2}n_{f,a-}+\frac{1}{2}n_{S,a-}(N_a+4)+\frac{1}{2}n_{A,a-}(N_a-4)=0
 \eea
where the sum is over all the representations ${\bf R_a}$ of  chiral multiplets.
 Solving these conditions at each node, one finds  for the first few choices of $\Z_n$ the series of
 anomaly-free quiver gauge theories listed in Table \ref{tabquiver}
{\tiny
\begin{center}
\begin{table}[ht]
\begin{tabular}{|c|c|}
\hline
\hline
 & $\Z_3$ \\
\hline
V & $Sp/O(N_0) \times U(N_1)$\\
$\Phi^I$  & $3 \left[ N_0 \bar N_1+\ft12 N_1(N_1\pm 1)\right]$\\
$I_a=0$  & $N_0=N_1\pm 4$\\
\hline\hline
 & $\Z_4$ \\
\hline
V & $Sp/O(N_0) \times Sp/O(N_2) \times U(N_1)$\\
$\Phi^I$  & $2(N_1 N_2 + N_0\bar N_1)+N_0  N_2+\ft12 N_1(N_1\pm 1)+\ft12 \bar N_1(\bar N_1\pm 1)$\\
$I_a=0$  & $N_0=N_2$\\
\hline\hline
 & $\Z_5$ \\
\hline
V & $Sp/O(N_0) \times U(N_1) \times U(N_2)$\\
$\Phi^I$   & $2  \left[ N_0\,  \bar N_1 +  N_1\,\bar  N_2 +
\ft12\,N_2 \, (N_2\pm 1) \right]+
 N_2\, N_0 + \bar N_1\, \bar N_2  + \ft12N_1(N_1\pm 1)$\\
$I_a=0$  & $N_0=N_1=N_2\pm 4$\\
\hline
\end{tabular}
\caption{\small Low energy theories arising from the quivers
described in the main text.} \label{tabquiver}
\end{table}
\end{center}
}
  We are interested in instanton generated superpotentials in these quiver gauge theories.
 A superpotential is generated by gauge instantons if and only if the number of fermionic zero
 modes in the instanton background ${\rm dim}\mathfrak{M}^{\rm ferm}_{a,k}$ and
 the $\beta$ function of the
 gauge theory $\beta_a $  satisfy the relation
 \cite{Bianchi:2007fx, Bianchi:2007wy}
 \be
  {\rm dim}\mathfrak{M}^{\rm ferm}_{k_a,N_a}=2 k_a\beta_a-4   \label{condw}
 \ee
 with $k_a$ the number of D(-1)-branes (instanton number) sitting on top of the $N_a$ D3-branes and\footnote{In our
conventions: ${\rm tr}_{\bf R} T^a \, T^b =\ell({\bf R}) \delta_{ab}$ and $\ell({\bf F})=\ft12$.  }
 \bea
 {\rm dim}\mathfrak{M}^{\rm ferm}_{k_a,N_a} &=& 2 k_a \left[ \ell({\bf Adj_a})+ \sum_{\bf R_a}  \ell({\bf R_a})\right] \nn\\
 &=& k_a\left[2 N_a+  n_{f,a+} + n_{S,a+}(N_a+2)+ n_{A,a+}(N_a-2) \right]\nn\\
\beta_{a}&=& 3\, \ell({\bf Adj_a})- \sum_{\bf R_a}  \ell({\bf R_a})\nn\\
&=&3 N_a - \ft12 n_{a,f+}  -\ft12 n_{S,a+} (N_a+2) -\ft12 n_{A,a+} (N_a-2) \label{betam}
 \eea
the number of fermionic zero modes around the instanton background.
  When condition (\ref{condw}) is satisfied an Affleck-Dine-Seiberg (ADS) like superpotential is generated
  \be
  W_{\rm non-pert}={\Lambda^{k_a\beta_a}\over \cI_{k_a\beta_a-3}}
  \ee
 with $\cI_{k_a \beta_a-3}$ denoting a gauge-invariant flavor-singlet made out of
  $k_a \beta_a -3$ chiral superfields.

   Non-perturbative D-brane instanton effects in the $\Z_3$ case
  have been studied in  \cite{Bianchi:2007fx, Bianchi:2007wy, Ibanez:2007tu}. A superpotential in this series is generated only for the U(4) gauge theory
   choice with $\Omega^-$-planes and $Sp(6)\times U(2)$ for
   $\Omega^+$-planes.    Here we are interested in quiver
    gauge theories with $U(5)$ gauge groups and GUT matter. Such matter contents can be
    realized by taking the orthogonal series and choosing properly the  $N_a$'s in table \ref{tabquiver}. In
 table \ref{tabmatter} we list the matter content of these theories.
\begin{table}[ht]
\begin{center}
\begin{tabular}{|c|c|c|c|}
\hline
subgroup & matter content & gauge group& $(N_0,N_1,..)$ \\
\hline
 $\Z_3$ & $3\,({\bf 10}+{\bf \bar 5})$ & $  U(5)$ & (1,5)\\
\hline
 $\Z_4$ & $({\bf 10}+{\bf \bar 10})+2({\bf 5}+{\bf \bar 5})+{\bf 1}$ & $  U(5)$ &(1,5,1)\\
 \hline
 $\Z_5$ & $2({\bf 10}+{\bf \bar 5})+({\bf 5}+{\bf \bar 5})+2\times{\bf 1}$ & $U(1)\times U(5)$ & (1,1,5)\\
  \hline
\end{tabular}
\end{center}
\caption{\small Matter content of the quiver theories with a
$U(5)$ factor in the gauge group.} \label{tabmatter}
\end{table}%
  Using (\ref{betam})  one can easily check that none
  of these theories satisfy the condition (\ref{condw}) and therefore
 there is apparently no room for instanton effects.  However, in presence of a
 tree level superpotential (\ref{wtree}), this is
 not the end of the story, since a ${\bf 5}-{\bf \bar 5}$ pair
 can get mass from  Yukawa interactions and decouples from the spectrum
  allowing for an instanton generated superpotential.
The focus on this paper is on the non-perturbative dynamics of
this $\Z_5$-quiver gauge theory.

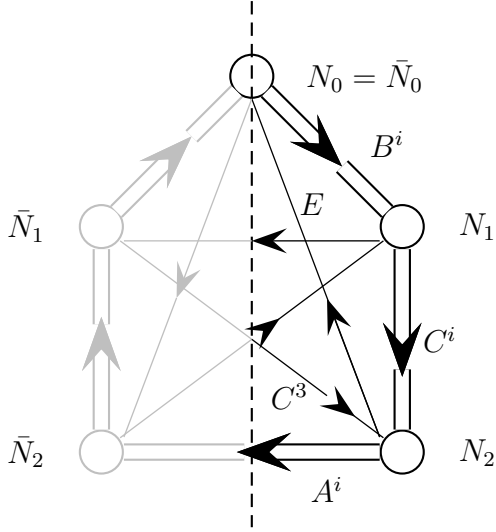
\begin{figure}[ht]
\label{fig}
\setlength{\unitlength}{2mm}
\begin{center}
\begin{picture}(50,25)
\pscircle[linecolor=lightgray](3,1){.3}
\pscircle(7,1){.3}
\rput(8,1){$N_2$}\rput(2,1){$\bar N_2$}
\pscircle[linecolor=lightgray](3,4){.3}
\pscircle(7,4){.3}
\rput(8,4){$N_1$}\rput(2,4){$\bar N_1$}
\pscircle(5,6){.3}
\rput(6.5,6){$N_0=\bar N_0$}\rput(6.8,5.1){$B^i$}\rput(7.5,2.5){$C^i$}\rput(6,.5){$A^i$}\rput(5.5,1.75){$C^3$}\rput(5.8,4.3){$E$}
\psline[doubleline=true,doublesep=5pt,linecolor=lightgray](3,2.7)(3,3.7)
\psline[doubleline=true,doublesep=5pt,linecolor=lightgray]{->}(3,1.3)(3,2.8)

\psline[doubleline=true,doublesep=5pt,linecolor=lightgray](3.3,1)(4.9,1)
\psline[doubleline=true,doublesep=5pt]{<-}(4.8,1)(6.7,1)
\psline[doubleline=true,doublesep=5pt](7,1.3)(7,2.1)
\psline[doubleline=true,doublesep=5pt]{<-}(7,2)(7,3.7)
\psline[doubleline=true,doublesep=5pt]{->}(5.2,5.8)(6.2,4.8)
\psline[doubleline=true,doublesep=5pt](6.2,4.8)(6.8,4.2)
\psline[doubleline=true,doublesep=5pt,linecolor=lightgray]{->}(3.2,4.2)(4.2,5.2)
\psline[doubleline=true,doublesep=5pt,linecolor=lightgray](4.2,5.2)(4.8,5.8)
\psline[linestyle=dashed](5,7)(5,0)
\psline[linewidth=.5pt](5,5.7)(6.7,1.2)\psline[linewidth=.5pt,arrowsize=6pt 7]{<-}(6,3.04)(6.7,1.2)
\psline[linewidth=.5pt,arrowsize=6pt 7]{>-}(5,2.5)(6.75,3.81)\psline[linecolor=lightgray,linewidth=.5pt](3.25,1.19)(5,2.5)
\psline[linewidth=.5pt,linecolor=lightgray,arrowsize=6pt 7]{->}(5,5.7)(4,3.05)\psline[linewidth=.5pt,linecolor=lightgray](4,3.05)(3.3,1.2)
\psline[linewidth=.5pt,linecolor=lightgray](3.3,3.81)(5,3.81)\psline[linewidth=.5pt,arrowsize=6pt 7]{<-}(5,3.81)(6.7,3.81)
\psline[linewidth=.5pt,linecolor=lightgray](3.25,3.81)(5,2.5)\psline[linewidth=.5pt](5,2.5)(6,1.75)
\psline[linewidth=.5pt,arrowsize=6pt 7]{>-}(6,1.75)(6.75,1.19)
\end{picture}
\caption{The $\Z_5$ quiver. The gauge group $U(5)=SU(5)\times U(1)_5$ ($U(1)_1$) is given by the stack with $N_2$ ($N_1$) branes respectively.
The figure is symmetric with respect to the orientifold plane which is represented as a dashed line.
Our convention is that the fields represented by entering arrows have negative quantum numbers.}
\end{center}
\end{figure}

    The field content of the $\Z_5$ quiver gauge theory is
    summarized in table \ref{matter}.
    $U(5)\sim SU(5)\times U(1)_5$ and $U(1)_1$ are gauge symmetries while $SU(2)$ is a global
    symmetry rotating the two generations.
We name the various chiral multiplets (the invariant blocks of
$\Phi^I$) as $\{ A^{i  }  ,B^{i  }  ,C^{i  }  ,C^3,E\}$.
\begin{table}[ht]
\begin{center}
\begin{tabular}{|c|c|c|c|}
\hline
fields & $SU(2)_{F}$ & $SU(5)$ & $U(1)_1\times U(1)_5$ \\
\hline
$A^{i  }  _{uv}$ & ${\bf 2}$  & ${\bf 10}$ & $(0,2)$ \\
$B^{i  }  $ & ${\bf 2}$  & ${\bf 1}$ & $(-1,0)$ \\
$C^{{i  }   u} $ & ${\bf 2}$  & ${\bf \bar 5}$ & $(1,-1)$ \\
  $C^{3 u}$ & ${\bf 1}$  & ${\bf \bar 5}$ & $(-1,-1)$ \\
   $E_{u}$ & ${\bf 1}$  & ${\bf  5}$ & $(0,1)$ \\
\hline
\end{tabular}
\end{center}
\caption{\small Chiral matter field content. Indices ${i  }  =1,2$, $u=1,..5$ run over the fundamentals of the $SU(2)$
flavor and $SU(5)$ gauge
groups. }
\label{matter}
\end{table}%
The prepotential  follows from (\ref{wtree}) by restricting $\Phi^I$ to their $\Z_5$-invariant components and
can be written as
 \be W_{\rm tree} =  C^{{i  }   u}\,  B_{{i  }  }\, E_{u} +
 C^{{i  }   u}\,  A_{{i  }   uv}\,  C^{3 v}   \label{wtree2} \ee
with $u,v=1,..5$, ${i  }  =1,2$.
The prepotential (\ref{wtree2}) and the quantum numbers of table \ref{matter} can also be deduced from the quiver diagram in figure \ref{fig}
\vskip 1.5cm

\section{Non-perturbative dynamics}

\subsection{SQCD: a short review }

 The studies of gauge theories at the non-perturbative level
 have a long history. The prototypical
case is that of SQCD, i.e. ${\cal N}=1$ $U(N_c)$ gauge theory
with $N_f$ flavors $Q^i$ and $\tilde Q_i$, $ i=1,..N_f$,  in the
fundamental and anti-fundamental representations of the gauge
group respectively
(see \cite{Amati:1988ft,Shifman:1995ua,Shifman:1999mv,Peskin:1997qi,Terning:2006bq} for reviews).
A superpotential is generated by instantons
only for $N_f=N_c-1$.
 For a generic $N_f$, the strong coupling dynamics  is   described in terms of
 composite fields:  the mesons
 $ M^i_j$ and (for $N_f\geq N_c$) baryons $B^{[i_1...i_{N_c}]}$, $\tilde  B^{[i_1...i_{N_c}]}$
 \be
 M^i_j=Q^i\tilde Q_j \quad\quad B^{[i_1...i_{N_c}]}=Q^{[i_1}....Q^{i_{N_c}]}
 \quad\quad \tilde B^{[i_1...i_{N_c}]}=\tilde Q^{[i_1}....\tilde Q^{i_{N_c}]}
 \ee
 The case $N_f= N_c$ is special in the sense that the $B,\tilde{B}$ are singlets
 and satisfy the (classical) algebraic relation
 \be
{\rm  det}\, M-B \tilde B=0   \label{mbb}
 \ee
  A similar relation between composites will be found  in our analysis of the $\Z_5$-quiver gauge theory
 later in this section. Before analyzing the $U(5)$ theory, it is convenient to recall how
 the non-perturbative dynamics manifests in this more standard SQCD setting.

  At the quantum level the relation (\ref{mbb}) is deformed to
   \be
{\rm  det}\, M-B \tilde B=\Lambda^{\beta}   \quad\quad \beta=2 N_c
 \ee
  with $\Lambda$ the SQCD scale.  Introducing a Lagrangian multiplier U for this constraint an effective superpotential
  for the theory can written as
  \be
  W_{\rm eff}=W_{\rm tree}+U( {\rm  det}\, M-B \tilde B-\Lambda^{2 N_c}   )  \label{weff}
  \ee
   with $W_{\rm tree}$ the tree level superpotential. This formula can be tested by  taking for the tree level
   superpotential a mass term for the $d^{\rm th}$ quark-antiquark pair
   \be
  W_{\rm tree}=m Q^d \tilde Q_d
\ee
 In the limit $m\to \infty$ the massive quarks decouple and one is left with SQCD with $N_f=N_c-1$ flavors where
 an ADS superpotential is expected. Indeed writing
 \be
 M = \left(
\begin{array}{ccc}
  \hat M_{ij}   & X_j  \\
  Y_i   &  Z
\end{array}
\right) \quad\quad i,j=1,...N_c-1
 \ee
 and solving the F-flatness conditions for the massive components
 \be
 {\partial W_{\rm eff}\over \partial X_j}
 ={\partial W_{\rm eff}\over \partial Y_i}
 ={\partial W_{\rm eff}\over \partial B}
 ={\partial W_{\rm eff}\over \partial \tilde B}
 ={\partial W_{\rm eff}\over \partial U}=0
 \ee
 one finds
 \be
 X_j=Y_i=B=\tilde B=0 \quad\quad Z={\Lambda^{2 N_c} \over {\rm det} \hat M}
 \ee
Plugging this into (\ref{weff}) one finds
\be
W_{\rm eff}={ m \Lambda^{2 N_c }\over  {\rm det} \hat M}={ \hat \Lambda^{2 N_c+1 }\over  {\rm det} \hat M}
\ee
 as expected for the ADS superpotential in $U(N_c)$ SQCD with $N_f=N_c-1$ flavors after the identification
$\hat \Lambda^{2 N_c+1}= m \Lambda^{2 N_c }$ between the scales of the two theories.
Remarkable, the non-perturbative superpotential for SQCD with $N_f=N_c-1$ origins from the quantum deformation
of the classical relation between baryons and mesons in the parent theory with $N_f=N_c$ !.
 In the following section we will find similiar results  for  the $U(5)$ chiral theory
  emerging from the unoriented $\Z_5$ quiver.

\subsection{$\Z_5$-quiver superpotential}

The first step in defining the gauge theory in the strong coupling
regime is to identify the right gauge invariant degrees of
freedom, `mesons' and `baryons'.
 The two $U(1)$ factors are anomalous and decouple anyway in the IR.
 Therefore
 we should consider $SU(5)$ rather than $U(5)$ invariants.
We notice that in the absence of $W_{\rm tree}$ the fields $C^{{i
} u} $ and   $C^{3u}$ can be grouped into a triplet that we will
denote by  $C^{I u} $. $SU(5)$ invariants will then be built out
of the elementary fields  $A^{i  }_{uv}$, $C^{I u} $,  $E_{u}$ and
will be labelled by the indices $I=1,2,3$ and ${i  }  =1,2$ in the
fundamental of $SU(3)_C$ and $SU(2)_A$ flavor groups respectively.
The complete list of composites reads \bea
X^I &=& C^{I u} E_{u}\nn\\
Y_{ I}^{i  }   &=&  \ft12 \epsilon_{IJK}\, A^{{i  }  }_{ u_  1  u_  2}\,C^{J   u_  1}\,    C^{K   u_  2}\nn\\
\tilde{Y}^{i j} &=& \ft14 \epsilon^{ u_  1... u_  5} A^{{i  }  }_{ u_  1  u_  2}\,A^{{{j}  }}_{  u_  3  u_  4}\, E_{ u_  5} \nn\\
Z^{{i  }   I} &=& \ft{1}{12}\epsilon^{ u_  1... u_  5} A^{{i  }  }_{ u_  1  u_  2}\,
A^{{{j}  }}_{ u_  3  u_  4}\,A_{ u_  5 v, {{j}  }  }\, C^{I v  }
\label{composites}\eea
One can check that $X,Y,\tilde Y,Z$ satisfy the algebraic relations
 \bea
&&  Y^{{i  }  }_{I}\, Z_{{i  }  }^I =0 \nn\\
&&   \epsilon_{IJK} \,X^I\, Z_{{i  }  }^J\, Z^{{i  }   K}+Y_{i I}     \, \tilde{Y}^{ij}  Z^{I }_j =0 \label{xyzrel}
 \eea
 Indices $i,j $ are raised and lowered with $\epsilon^{i  j  },\epsilon_{ij }$.
 Together with $B_i$ one finds
 \be
 18=42_{\rm fields} -24_{\rm gauge}=20_{\rm composites}-2_{\rm relations}
 \ee
 gauge invariant degrees of freedom which one can use as coordinates
on the moduli space. As indicated, the very same number of degrees
of freedom can also be obtained by subtracting
 to the 42 degrees of freedom coming from elementary fields the
 $24={\rm dim}[{SU(5)}]$ independent generators of gauge invariance.

  It is important to notice that  the two relations in (\ref{xyzrel}) have engineering dimension 7 and 10 and therefore
  only the second one can get modified by a  non-perturbative $\Lambda^\beta=\Lambda^{10}$ term.
   Introducing the Lagrange  multipliers U, V for the two relations,
   the effective superpotential of the unoriented $\Z_5$-quiver
   gauge theory can then be written as
 \be
 W_{\rm eff}=  B_{{i  }  }  \, X^{i  }  +\delta^I_{ {i  }   }\,  Y^{{i  }   }_{ I}+V\,  Y^{{i  }  }_{I} \, Z_{{i  }  }^I+U\,(  Y_{ I}^{i  }     \, \tilde{Y}_{ i j}  Z^{I j  }+ \epsilon_{IJK} X^I\, Z_{{i  }  }^J\, Z^{{i  }   K}-\Lambda^{10})\label{weff0}
\ee
  Notice that the first two terms coming from $W_{\rm tree}$ break explicitly
  the `fake' $SU(2)\times SU(3)$ global symmetry  to its
  diagonal subgroup.

  The F-terms can be written as $F_s={\partial W_{\rm eff}\over \partial \Phi_s}$ with  $\Phi_s=(B,X,Y,\tilde Y,Z,U,V)$.
 Explicitly
  \bea
 F_{B_{i  }  } &=& X^{i  }   \nn\\
F_{X^I} &=&    B_{i  }   \delta^{i  }  _I +  U\,   \epsilon_{IJK}  \, Z_{{i  }  }^J\, Z^{{i  }   K}  \nn\\
F_{Y_{I}^{i  }  }&=&   \, \delta_{{i  }  }^I     + V\, Z_{{i  }  }^I  + U \,\tilde{Y}_{ i j  }  \,Z^{j  I} \nn\\
F_{\tilde{Y}^{i j }} &=& U \, Y_{ I ( i    }    \,Z_{ j  )}^I  \nn\\
F_{Z_{{i  }  }^I} &=&V\,  Y^{{i  }  }_{I} \,   +U\,( Y_{ I j  }  \, \tilde{Y}^{ij  }   + 2 \,
\epsilon_{IJK} X^K \, Z^{i   J} ) \nn\\
F_U&=&  Y_{ I {i  }   }  \, \tilde{Y}^{i  j}  Z_{ j }^I +\epsilon_{IJK} \,X^I\, Z_{{i  }  }^J\, Z^{{i  }   K}-\Lambda^{10}\nn\\
F_V&=&  Y^{{i  }  }_{I}\, Z_{{i  }  }^I  \label{fterm}
 \eea
We will momentarily discuss the solution of the above F-term
equations and then take into account the effect of  FI terms.

 \subsection{The supersymmetric vacuum}

 There is a one-parameter solution to the F-flatness conditions $F_s=0$ with $F_s$ given in (\ref{fterm}).
 This is given by setting all composite fields to zero except for
 \be
 Z^{21}=-Z^{12}   \quad\quad ~~~~~~~X^3={\Lambda^{10}\over 2 (Z^{12})^2} \quad\quad ~~~~~V={1\over Z^{12}}
 \ee
 Supersymmetric vacua are parametrized by the single modulus $Z^{12}$ .
 The superpotential $W_{\rm eff}$ vanishes along the valley of minima.

  \subsection{Dynamical supersymmetry breaking}

  Now let us consider the effect of turning on FI terms.
  In string theory this can be
  achieved by
  giving non-trivial VEV's to the two
  scalar fields (blowing up modes) belonging to
  the closed-string twisted sectors and thus living at the
  $\Z_5$-singularity. They are partners of the two axions acting as St\"uckelberg fields for
the anomalous $U(1)_1$ and $U(1)_5$ gauge bosons\footnote{The
precise
  combinations of twisted scalars entering the definition of the
   FI terms $\xi_{1,2}$ is determined by disk amplitudes with insertion of a closed-string
  scalar vertex in the bulk
  and an
  open-string auxiliary D-term vertex on the boundary. A similar computation determines the
  axion mixing with the anomalous $U(1)$ gauge bosons
  \cite{Angelantonj:1996uy, Aldazabal:2000sa, Anastasopoulos:2006cz}}.
   The $U(1)\times U(5)$ D-terms can be written as
  \bea
  D &=&  -|B_{i  }|^2 + |C^{\alpha u}|^2 -
|C^{3 u} |^2 + \xi_1 \nn\\
D^u_v &=& \bar A_{i  }^{u w} \, A^{{i}  }_{w v} - \bar C_{{{i}  } v} C^{{{i}  } u} -\bar C_{3 v} \, C^{3 u} +\bar E^u \, E_v  +\xi_2
\label{Dflat}
\eea
The F-terms following from $W_{\rm tree}$ can be written as
    \bea
    F_{B_i} &=& C^{iu}E_u \nn\\
    F_{A_i} &=& C^{iu}C^{3v} \nn\\
    F_{C^3} &=& A_{iuv}C^{iu} \nn\\
    F_{C^i} &=& B_i E_u+A_{iuv}C^{3v}
    \eea
   For $\xi_2=0$, $\xi_1>0$, there is an  SU(5) preserving solution
   to both D and F-flatness conditions given by taking
  \be
 \langle B_{i } \rangle=(m,0) \quad\quad~~~~~~~      \xi_1=m^2
  \ee
    and setting all remaining fields $A^i,C^I,E$ to zero. This defines a perturbative supersymmetric vacuum
    of the theory.
   We notice that for $m\neq 0$ the superpotential interactions result into a mass term for
   $C^{1u}$ and $E^u$
   via the coupling
  \be
  W_{\rm tree}=m C^{1u}E_u+ \ldots
  \ee
  In the limit $m\to \infty$  these fields decouple  from the spectrum and  we find a U(5) gauge theory
 with two generations of $({\bf 10}+{\bf \bar 5})$. The condition
 (\ref{condw}) is now satisfied with  $\beta = 11$
  and ${\rm dim}\mathfrak{M}^{\rm ferm}_{k=1}=18$ and therefore
 a superpotential is generated by gauge instantons. The form of the superpotential is completely
 fixed by dimensional analysis and $SU(5)\times SU(2)$ invariance. Indeed there is a unique
 $SU(5)\times SU(2)$ invariant of dimension eight:  $Z^{i \alpha}Z_{i \alpha}$. The ADS superpotential
  can then be written as
 \be
 W_{\rm ADS}={\hat{ \Lambda}^{11} \over Z^{i \alpha}Z_{i \alpha}}
 \ee
  with $\alpha=2,3$ labelling the massless $C^{\alpha u}$ components and $\hat{\Lambda}^{11}=m\Lambda^{10}$ the effective scale of the massless theory. The effective superpotential can be
  written as the sum of $W_{\rm ADS}$ and $W_{\rm tree}$ with all massive fields set to zero.
  \be
  W_{\rm eff}=A^1_{uv} \,C^{2u} \, C^{3v}+{m\,\Lambda^{10}\over Z^{i \alpha}Z_{i \alpha}} \label{resultw}
  \ee
      Summarizing, for $m>>\Lambda$ the unoriented $\Z_5$
      quiver GUT-like theory is effectively described  in terms of
      an $SU(5)$ gauge theory with two generations of
      matter $({\bf 10}+{\bf \bar 5})$
       and the superpotential (\ref{resultw}).

       Similarly to SQCD, in which the $N_f=N_c-1$ case is
       recovered from that with $N_f=N_c$ by
decoupling a massive flavor, (\ref{resultw}) can be derived from
the effective superpotential (\ref{weff0}) after solving the
F-flatness conditions for the massive fields in favor of the
massless ones.
       The gauge invariant massless degrees of freedom are encoded in the following composites
        \bea
        Y_{ 1}^{i  }   &=&  A^{{i  }  }_{ u_  1  u_  2}\,C^{\alpha   u_  1}\,    C_{\alpha}^{   u_  2}\nn\\
        Z^{i \alpha}  &=& \epsilon^{ u_  1... u_  5} A^{{i  }  }_{ u_  1  u_  2}\,
A^j_{ u_  3  u_  4}\,A_{ u_  5 j,k  }\, C^{\alpha v  }
        \eea
         and the singlets $B_i$. Altogether they provide us with the expected $8=32-24$  degrees of freedom.
     Solving the F-flatness conditions for the
     remaining composites one finds
     \bea
     X^1 &=& {\Lambda^{10}\over Z^{i \alpha}Z_{i \alpha}}\quad\quad X^2 = {Y_1^2 \over m}    \quad\quad U=-{m\over Z^{i \alpha}Z_{i \alpha}}
      \quad\quad V={Z^{23} \over Z^{i \alpha}Z_{i \alpha}}\nn\\
        \tilde Y_{22}& =&  -{2Z^{13}\over m}    \quad\quad  ~~~~   \tilde Y_{12} =  {Z^{23}\over m}   \quad\quad ~~~B_i=(m,0)
      \label{solutionb}
       \eea
 with all remaining fields set to zero. Plugging (\ref{solutionb})
 into (\ref{weff0}) one finds
 \be
 W_{\rm eff}=Y_1^1+{m \Lambda^{10}\over Z^{i \alpha}Z_{i \alpha}}
\label{potcomposites} \ee
 in perfect agreement with (\ref{resultw}).

 It is easy to check that the F-flatness conditions for the
 massless fields cannot
 all be simultaneously satisfied and therefore {\it supersymmetry
 is dynamically broken} by instanton effects \cite{Affleck:1984mf,Affleck:1984xz,Amati:1988ft}.
The $U(5)$ gauge theory under consideration here with
superpotential (\ref{resultw})
 is indeed one of the classic examples of dynamically supersymmetry
 breaking. Nicely
 string theory provides a
 precise realization of this theory where the superpotential is not
 an {\it ad hoc} choice
 but rather a result of string dynamics.
  The fact that supersymmetry is broken can be seen
  alternatively by considering the Konishi anomaly equation
 \be
 {1\over 4} \bar{D}^2(\Phi^r e^{gV} \Phi^\dagger_s)=  \Phi^r {\partial W_{tree} \over \partial\Phi^s} + {g^2\over
 32\pi^2}  \delta^r{}_s tr_{R_s}(W^\alpha W_\alpha)
 \label{konanomaly}\ee
  valid for any $r$ and $s$.
  Using the fact that $\bar D$ annihilate supersymmetric states the
  VEV of the leftt hand side of this equation on a supersymmetric
  vacuum is zero.
  Taking the diagonal term in (\ref{konanomaly}) (no sum over $s$) and the lowest component of the superfields, 
we get
  \be
\langle    \phi^s {\partial W_{tree} \over \partial \phi^s}\rangle
+ {g^2\over
 32\pi^2} \,{\rm Tr}_{R_s} \langle \lambda \lambda \rangle  =0  \label{konishi}
  \ee
  The first term cancels for $\phi_s=A^2$ since this field is absent
  from the tree
  level superpotential (\ref{resultw}). The
  presence of a gaugino condensate ${\rm Tr}_{R_s}\langle \lambda \lambda \rangle \neq 0$
  then implies that supersymmetry is broken.

\subsection{The weak coupling description}

The results presented in the previous subsection
 can be further supported by a weak coupling
analysis. To this purpose we notice that the $SU(5)$
$\beta$-function and the number of fermionic zero
modes\footnote{This number is $2\ell ({\bf R})$ for each fermion in the ${\bf R}$
representation.}
 (i.e. the dimension of the fermionic moduli space) for the $k=1$
 gauge instanton,
are given by (\ref{betam})
\bea
 {\rm dim}\mathfrak{M}^{\rm ferm}_{k=1} &=& 10_{\lambda}+  3_{\psi_C}+1_{\psi_E} +6_{\psi_A}=20 \nn\\
\beta&=& 10 \label{betam2}
 \eea
 where we have indicated in the first line the origin of the fermionic zero modes.
${\rm dim}\mathfrak{M}^{\rm ferm}_{k=1}$ and $\beta$ do not satisfy the relation (\ref{condw})
 and therefore a superpotential of the ADS type is not generated by instantons in this case.
Nonetheless an instanton dominated correlator is easily identified
to be
 \be
 \langle X^I\, Z^{iJ} \,Z^{jK}  \rangle =\epsilon^{IJK}\,
 \epsilon^{ij} \,\Lambda^{10} \label{correl}
 \ee
Indeed each scalar soaks up a
gaugino and a matter fermion zero mode, $X Z^2 \sim E C^3 A^6$ is
precisely what is needed in order to soak up the fermionic zero
modes in (\ref{betam}). Here we use the gauge invariant
combinations $X^I$, $Z^{iJ}$, defined in (\ref{composites}), only
for the sake of book-keeping. The correlator (\ref{correl}) can be
computed in the weak coupling regime, where the relevant degrees
of freedom are the elementary fields $A,C,E$ and $X,Z$ are
quadratic and quartic gauge invariant combinations of these
variables.

Using the Konishi identity
(\ref{konishi}) with $\Phi^s=E$ and taking $B_i=(m,0)$ one finds
\be {g^2\over
 32\pi^2}  \langle {\rm tr}\, \lambda \lambda \, Z^{i\alpha} \,Z_{i\alpha}\rangle = m\,   \Lambda^{10}
\ee
 in agreement with (\ref{solutionb}) and (\ref{potcomposites}).
Alternatively this result can be justified by noticing that an
insertion of $\lambda\lambda$ soaks up two gaugino zero modes,
$Z^2$  16 fermionic zero modes, and the Yukawa interaction  $B
\psi_C \psi_E$  the last 2 fermionic zero modes.

  \section{Anomaly matching conditions}

  In this section we test the strong coupling description of the $\Z_5$-quiver gauge theory by showing that
  the anomalies of the global symmetries computed in the `microscopic' theory match those of the `macroscopic'
  theory
  described in terms of baryons and mesons. Omitting
   the $SU(5)$ singlet fields $B_i$ fields, which contribute to the
strong and weak coupling phases in an identical way, and
neglecting the Yukawa couplings for the time being, the global
symmetry of the theory turns out to be $SU(3)_C\times SU(2)_A\times U(1)_{A}\times
U(1)_{C}\times U(1)_{E}$ with the subscripts indicating the fields
with unit charge which the corresponding $U(1)$ acts on. We now
consider a particular solution of (\ref{xyzrel}) preserving the
largest possible symmetry \be
 X^3=\Lambda^{2}   \quad\quad  Z^{iI}=\frac{1}{\sqrt{2}}\Lambda^{4}\epsilon^{i I}  \quad\quad I=1,2 \label{solcons}
\ee
 with all remaining composite fields set to zero. The symmetry preserved by this solution is
 $G_F =SU(2)\times U(1)_M\times U(1)_N$ where the two U(1)'s are defined in such a way that
 fields appearing in the solution (\ref{solcons}) are uncharged .
 Table \ref{chargesSU2U12}  summarizes the  charges of
 elementary and composite fields under this symmetry.
  The $SU(2)$ part is the diagonal
subgroup of the $SU(3)_C\times SU(2)_A$ global symmetry as follows
from the invariance of (\ref{solcons}).
 \begin{table}[ht]
\begin{center}
\begin{tabular}{|c|cccc|ccccccccc|}
\hline
fields &  $A^{i  }_{uv}$  & $C^{i u}  $& $C^{3 u} $ &   $E_{u}$  & $X^3$ & $X^i$  & $Y^i_3$ & $Y^i_j$ & $\tilde{Y}^{(ij)} $ & $Z^{i3}$ & $Z^{ij}$  & U&V\\
\hline
     $SU(2)$ &     ${\bf 2}$  &  ${\bf 2}$  & ${\bf 1}$ & ${\bf 1}$ &               ${\bf 1}$  &  ${\bf 2}$  & ${\bf 2}$ & ${\bf 3},{\bf 1}$ &  ${\bf 3}$ &
     ${\bf 2}$ & ${\bf 3},{\bf 1}$ &  ${\bf 1}$& ${\bf 1}$\\
     $U(1)_M$  &     1& -3 & 0 & 0 &               0& -3&-5&-2& 2&3& 0&0&2\\
      $U(1)_N$ & 0 & 0& 1& -1             &             0&-1&0&1&-1&1&0&0&-1\\
  \hline
  $SU(5)$ &     ${\bf 10}$  &  ${\bf \bar 5}$  & ${\bf \bar 5}$ & ${\bf 5}$ &               ${\bf 1}$  &  ${\bf 1}$  & ${\bf 1}$ & ${\bf 1}$ &  ${\bf 1}$ &
     ${\bf 1}$ & ${\bf 1}$ &  ${\bf 1}$& ${\bf 1}$\\
     \hline
\end{tabular}
\end{center}
\caption{\small  $SU(2)\times U(1)^2$ charges of elementary and composites fields. The last row displays the multiplicities
of the given field}
\label{chargesSU2U12}
\end{table}%
One can easily check that mixed $U(1)_{M,N} SU(5)^2$ anomalies
cancel  i.e. $U(1)_{M,N}$ are truly conserved currents of the
theory. Indeed using \be I_{U(1)_{a} SU(5)^2} = \sum_{s} {\rm
dim}({\bf R}^s_{SU(2)}) \, Q_{a}(\Phi_s) \ell({\bf R}^s_{SU(5)})
\quad\quad  a=M,N \ee where $s$ runs over all elementary fields in
table \ref{chargesSU2U12}, one finds \bea
I_{U(1)_{M} SU(5)^2} &=& 2\ft32+2\,\ft12(-3)=0\nn\\
I_{U(1)_{N} SU(5)^2} &=& \ft12(1-1)=0 \eea Now let us consider
cubic anomalies of the global currents. Since $SU(2)$ admits no
complex representation there is no cubic anomaly made only of
$SU(2)$ currents\footnote{Including $B^i$ the number of doublets
is even and no global anomaly appears.}. The remaining cubic
anomalies can be written as \bea
I_{U(1)_M^n U(1)_N^{3-n} } &=& \sum_{s}  {\rm dim}({\bf R}^s_{SU(2)\times SU(5)}) \, \,Q_M(\Phi_s)^n \,Q_N(\Phi_s)^{3-n}  \nn\\
I_{U(1)_{a} SU(2)^2} &=& \sum_{s}  {\rm dim}({\bf R}^s_{SU(5)}) \,
Q_{a}(\Phi_s) \ell({\bf R}^s_{SU(2)}   ) \quad\quad a=M,N
\eea
The spectrum of $U(1)_N$ charges is non-chiral for elementary
fields and therefore all anomalies involving $U(1)_N$ cancel. This
is in agreement with the cubic anomalies computed in
    the strong coupling regime where the theory is described
    in terms of the baryons/mesons $X,Y,\tilde{Y},Z$
\bea
I^{XYZ}_{U(1)_N^3}&=&(-1)^3 (2+3+1)+(1)^3(4+2)=0  \nn\\
I^{XYZ}_{U(1)_N^2 U(1)_M}&=& 2(-3)+4(-2)+3(2)+2(3)+(2)=0  \nn\\
I^{XYZ}_{U(1)_N U(1)_M^2}&=& 2(-9)+4(4)+3(-4)+2(9)+(-4)=0  \nn\\
I^{XYZ}_{U(1)_N SU(2)^2}&=& \ft12(-1+1)+2(1-1)=0 \eea On the other
hand the non-trivial anomalies in the microscopic theory are
  \bea
I^{ACE}_{U(1)_M^3} &=& 20\,(1)^3-10 (-3)^3=-250\nn\\
I^{ACE}_{U(1)_M SU(2)^2} &=& 10\ft12-15\ft12=-\ft52 \eea matching
again those in the macroscopic theory
 \bea
I^{XYZ}_{U(1)_M^3}&=&2(-3)^3+2(-5)^3+4(-2)^3+3(2)^3+4(3)^3+2=-250
\nn\\
I^{XYZ}_{U(1)_M SU(2)^2}&=&\ft12(-3-5+3)+(2-2)=-\ft52
 \eea
  The perfect match between the anomalies in the two regimes
  provides a robust consistency test of the proposed non-perturbative description of the $\Z_5$-quiver gauge theory.

\section{D-instanton description}

In this section we briefly describe the D-instanton derivation of
the non-perturbative effects relevant to our analysis of the
$\Z_5$ quiver theory.

Non-perturbative effects, generated by D-brane instantons, have
been the subject of a dedicated effort of several groups in the
past couple of years. The emerging picture suggests that there are
two kinds of D-brane instantons. The first class, termed ``gauge''
instantons correspond to Euclidean D-branes wrapping the same
cycle as a stack of ``physical'' branes present in the background
\cite{Akerblom:2006hx,Argurio:2007vqa,Bianchi:2007wy}. The second
class, termed ``exotic'' or ``stringy'' instantons, correspond to
Euclidean D-branes wrapping a cycle not wrapped by any
``physical'' branes realizing the gauge theory
\cite{Blumenhagen:2006xt} \nocite{Ibanez:2006da,Bianchi:2007fx,
Argurio:2007vqa,Bianchi:2007wy}-\cite{Ibanez:2007rs}. Obviously,
in a quiver gauge theory, what looks as a `gauge' instanton for a
given gauge group may look as an `exotic' instanton for a
different gauge group factor.

In the unoriented $\Z_5$ quiver theory, the relevant
D(-1)-instanton is an $k=1$ instanton sitting on the $U(5)$ node
of the quiver. Being a gauge instanton, the supermoduli space will
comprise both neutral D(-1)D(-1) and charged D(-1)D3 zero-modes.

 The field content of the D3D(-1) gauge theory describing the
 dynamics of the D-instanton is summarized in table \ref{fzminst}.
 States are organized according to their charges under the
 $U(1)_1\times U(1)_5 \in U(1)\times U(5)$ gauge symmetries,
 the $SU(2)$ flavor group and $U(1)_{\rm inst}$ D(-1)-symmetry.
 The various indices run over the following domains
 $\alpha,\dot\alpha=1,2$ (Lorentz spinors), $i=1,2 $ (flavor) , $u=1,..5$ (gauge fundamentals).
 This content follows from that of $k_2=k_3=1$
 fractional D(-1) and $N_2=N_3=5$, $N_0=N_1=N_3=1$
 fractional D3-brane system in flat space after keeping
 the components invariant under
 $\Z_5$ and $\Omega$.

 \begin{table}[ht]
\begin{center}
\begin{tabular}{|c|c|c|c|c|}
\hline
origin & fields & $  SU(2)_{\rm F}$ & $SU(5)$ & $U(1)_{1}\times U(1)_{5}\times U(1)_{\rm inst}$ \\
\hline
D3D3 & $A^{i}  _{uv}$ & ${\bf 2}$  & ${\bf 10}$ & $(0,2,0)$ \\
 &$B^i  $ & ${\bf 2}$  & ${\bf 1}$ & $(-1,0,0)$ \\
 &$C^{i   u} $ & ${\bf 2}$  & ${\bf \bar 5}$ & $(1,-1,0)$ \\
&  $C^{3 u}$ & ${\bf 1}$  & ${\bf \bar 5}$ & $(-1,-1,0)$ \\
   &$E_{u}$ & ${\bf 1}$  & ${\bf  5}$ & $(0,1,0)$ \\
\hline \hline
D(-1)D(-1) & $(a_{\alpha\dot \alpha} ,\Theta_\alpha^{0} ,\bar\Theta^{\dot\alpha}_{0}) $  & $ {\bf 1}$  & ${\bf 1}$ & $(0,0,0)$ \\
 & $(\bar\chi_i, \bar\Theta^{\dot\alpha}_i) $ & $ {\bf 2}$  & ${\bf 1}$ & $(0,0,-2)$ \\
 & $\chi^i$ & $ {\bf 2}$  & ${\bf 1}$ & $(0,0,2)$ \\
D(-1)D3 & $( \bar{w}^u_{\dot\a},\bar{\nu}^{0 u})$ & $ {\bf 1}$  & ${\bf 5}$ & $(0,1,-1)$ \\
& $( {w}_u^{\dot\a}, \nu_u^{0}) $ & $ {\bf 1}$  & ${\bf 5}^*$ & $(0,-1,1)$ \\
&$\nu^{i u}$ & $  {\bf 2}$  & ${\bf 5}$ & $(0,1,1)$ \\
&$\bar{\nu}^i$ & $  {\bf 2}$  & ${\bf 1}$ & $(1,0,-1)$ \\
&$\bar{\nu}^{3}$ & $ {\bf 1}$  & ${\bf 1}$ & $(-1,0,-1)$ \\
& $\nu^{3}$ & $ {\bf 1}$  & ${\bf 1}$ & $(0,0,1)$ \\
\hline\hline
& $d\mathfrak{M}$ & ${\bf 1}$  & ${\bf 1}$ & $(-1,-10,0)$ \\
\hline
\end{tabular}
\end{center}
\caption{\small Chiral fields and supermoduli } \label{fzminst}
\end{table}%

The superpotential
is defined by the following integral over the instanton moduli
space \be \int d^4 x d^2 \Theta \, W(\Phi_s)= \Lambda^{10} \int
d\mathfrak{M}  e^{-\cS_B-\cS_F } \ee
with
 \bea \cS_F
&=& \bar\Theta_{0 \dot\a}( w^{\dot\a}_u \bar\nu^{0u} + \nu^0_u
\bar{w}^{\dot\a u}) + \chi^i  \bar\nu^3 \bar\nu_i + \bar\chi_i
\nu^0_u \nu^{u,i } \nn \\ &&+ \overline{A}_{i  uv} \nu^{u,i }
\bar\nu^{0v} + \overline{E}_u \nu^3 \bar\nu^{0u} + \overline{C}_3
\nu^0_u \bar\nu^3 + \overline{C}^u_i  \nu^0_u \bar\nu^i  \\ &&+
B_{i} \nu^3 \bar\nu^i  + C^3_u \nu^{ui } \bar\nu_{i}
+ C^i_u \nu^{u i } \bar\nu_{3}\label{yuk}\\
 \cS_B &=& \overline\chi_i \chi^i
\overline{w}^u_{\dot\a}{w}_u^{\dot\a} + \chi^i
\overline{w}^u_{\dot\a}\overline{w}^{v\dot\a} \overline{A}_{iuv} +
\overline\chi_i
{w}_u^{\dot\a}{w}_{v\dot\a} A^{iuv} \nn \\
&&+ {w}_u^{\dot\a}\overline{w}^v_{\dot\a}[\overline{A}_{ivw}A^{iwu}
+ \overline{E}_v {E}^u + {C}^3_v \overline{C}^{ \, u}_3 + {C}^i_v
\overline{C}^u_i ] \eea
The fermionic zero-modes, except for
$\Theta^0$,  are lifted by Yukawa-type interactions in (\ref{yuk}).
 Roughly the fermionic integrals bring down 9 scalar fields in the numerator
  such that the $18={\rm dim} \mathfrak{M}^{\rm ferm}_{U(5),k=1}-2$ fermionic zero modes besides $\Theta^{0\alpha}$ are soaked up.
  Bosonic gaussian integrals should bring scalar fields in the denominator in such a way that the
  resulting expression for $W(\Phi_f)$ is holomorphic as expected.
Although the structure of the integrand supports our previous
derivation of the non-perturbative superpotential, the explicit
evaluation of the integral is rather involved and goes beyond the
scope of this work.

\section{Summary of results}

In this paper we have studied the non-pertubative dynamics of an
unoriented $\Z_5$ quiver gauge theory  with GUT-like gauge group
 $U(5)\times U(1)$, chiral matter in the $(2\times {\bf 10}+3\times {\bf \bar 5}+{\bf 5}+2\times {\bf 1})$  representations and a cubic superpotential.
At strong coupling, the dynamics of the gauge theory is described
by an effective superpotential given in terms of a  set of gauge
invariant variables, the baryons and mesons,  satisfying a quantum
deformed constraint.
  The gauge theory has two distinct phases depending on whether
 FI terms are turned on or not. In absence of FI terms the effective superpotential admits a line of supersymmetric vacua.
Turning  on a FI term for the $U(1)_1 \not\in U(5)$ the theory
undergoes a Higgs phase where an SU(5) singlet  gets a vacuum
expectation value and a $({\bf 5}+{\bf \bar 5})$ pair gets mass
from Yukawa interactions. The resulting    $U(5)$ gauge theory
with two generations of $({\bf 10}+{\bf \bar 5})$  is one of the
classical examples
 of chiral gauge theories with dynamical supersymmetry breaking via
 gaugino condensation \cite{Amati:1988ft,Affleck:1984mf,Affleck:1984xz}.
 We remark that supersymmetry is dynamically broken in this theory only for a bizarre choice of the tree level superpotential
 that breaks the $SU(2)$ flavor symmetry  rotating the two generations.
 Remarkable, this is a characteristic feature of the superpotential resulting from $\Z_5$ quiver
 in the Higgs phase.  The proposed strong coupling dynamics is tested
 by matching the anomalies of the effective theory described
in terms of the baryon/meson composites with those computed in terms of elementary
 fields.

Finally we have sketched how to derive the non-perturbative
superpotential as a D-instanton effect. We stress that
non-perturbative effects considered here origin only from  gauge
instantons i.e. fractional D(-1)-branes sitting at the $U(5)$ node
of the quiver. Indeed, in the quiver under consideration  there is
no room for exotic D-brane instantons i.e. those D(-1) branes sitting on top of nodes with no D3 branes.
 An alternative scenario of supersymmetry breaking based on exotic instanton interactions in presence
 of a $G_{(0,3)}$-flux was recently studied  in \cite{Billo':2008sp,Billo':2008pg}.

\vskip 1cm \noindent {\large {\bf Acknowledgments}} \vskip 0.2cm
The authors would like to thank S.~Ferrara, R.~Poghossyan,
G.~C.~Rossi, M.~Samsonyan and E.~Witten for interesting
discussions. One of us (M.~B.) would like to thank KITP in Santa
Barbara and GGI in Arcetri (Florence) for hospitality during
completion of this project. This work was partially supported by
the European Commission FP6 Programme MRTN-CT-2004-512194
``{Superstring Theory}'' and MRTN-CT-2004-503369 ``{The Quest for
Unification: Theory Confronts Experiment}'', by the Italian
MIUR-PRIN contract 20075ATT78 and by the NATO grant
PST.CLG.978785.
\noindent \vskip
1cm

\newpage

\begin{thebibliography}{10}

\bibitem{Witten:1981nf}
E.~Witten, \emph{{Dynamical Breaking of Supersymmetry}},
\href{http://dx.doi.org/10.1016/0550-3213(81)90006-7}{Nucl. Phys. {\bf B188}
  (1981)  513}.

\bibitem{Witten:1982df}
E.~Witten, \emph{{Constraints on Supersymmetry Breaking}},
\href{http://dx.doi.org/10.1016/0550-3213(82)90071-2}{Nucl. Phys. {\bf B202}
  (1982)  253}.

\bibitem{Amati:1988ft}
D.~Amati, K.~Konishi, Y.~Meurice, G.~C. Rossi, and G.~Veneziano,
  \emph{{Nonperturbative Aspects in Supersymmetric Gauge Theories}},
\href{http://dx.doi.org/10.1016/0370-1573(88)90182-2}{Phys. Rept. {\bf 162}
  (1988)  169--248}.

\bibitem{Affleck:1984mf}
I.~Affleck, M.~Dine, and N.~Seiberg, \emph{{Exponential hierarchy from
  dynamical supersymmetry breaking}},
\href{http://dx.doi.org/10.1016/0370-2693(84)91047-5}{Phys. Lett. {\bf B140}
  (1984)  59}.

\bibitem{Affleck:1984xz}
I.~Affleck, M.~Dine, and N.~Seiberg, \emph{{Dynamical Supersymmetry Breaking in
  Four-Dimensions and Its Phenomenological Implications}},
\href{http://dx.doi.org/10.1016/0550-3213(85)90408-0}{Nucl. Phys. {\bf B256}
  (1985)  557}.

\bibitem{Berenstein:2005xa}
D.~Berenstein, C.~P. Herzog, P.~Ouyang, and S.~Pinansky, \emph{{Supersymmetry
  Breaking from a Calabi-Yau Singularity}}, JHEP {\bf 09} (2005)  084,
\href{http://arxiv.org/abs/hep-th/0505029}{{\tt arXiv:hep-th/0505029}}.

\bibitem{Franco:2005zu}
S.~Franco, A.~Hanany, F.~Saad, and A.~M. Uranga, \emph{{Fractional branes and
  dynamical supersymmetry breaking}}, JHEP {\bf 01} (2006)  011,
\href{http://arxiv.org/abs/hep-th/0505040}{{\tt arXiv:hep-th/0505040}}.

\bibitem{Bertolini:2005di}
M.~Bertolini, F.~Bigazzi, and A.~L. Cotrone, \emph{{Supersymmetry breaking at
  the end of a cascade of Seiberg dualities}},
  \href{http://dx.doi.org/10.1103/PhysRevD.72.061902}{Phys. Rev. {\bf D72}
  (2005)  061902},
\href{http://arxiv.org/abs/hep-th/0505055}{{\tt arXiv:hep-th/0505055}}.

\bibitem{Diaconescu:2005pc}
D.-E. Diaconescu, B.~Florea, S.~Kachru, and P.~Svrcek, \emph{{Gauge - mediated
  supersymmetry breaking in string compactifications}}, JHEP {\bf 02} (2006)
  020,
\href{http://arxiv.org/abs/hep-th/0512170}{{\tt arXiv:hep-th/0512170}}.

\bibitem{Florea:2006si}
B.~Florea, S.~Kachru, J.~McGreevy, and N.~Saulina, \emph{{Stringy instantons
  and quiver gauge theories}}, JHEP {\bf 05} (2007)  024,
\href{http://arxiv.org/abs/hep-th/0610003}{{\tt arXiv:hep-th/0610003}}.

\bibitem{Aharony:2007db}
O.~Aharony, S.~Kachru, and E.~Silverstein, \emph{{Simple Stringy Dynamical SUSY
  Breaking}}, \href{http://dx.doi.org/10.1103/PhysRevD.76.126009}{Phys. Rev.
  {\bf D76} (2007)  126009},
\href{http://arxiv.org/abs/0708.0493}{{\tt arXiv:0708.0493 [hep-th]}}.

\bibitem{Buican:2008qe}
M.~Buican and S.~Franco, \emph{{SUSY breaking mediation by D-brane
  instantons}}, \href{http://dx.doi.org/10.1088/1126-6708/2008/12/030}{JHEP
  {\bf 12} (2008)  030},
\href{http://arxiv.org/abs/0806.1964}{{\tt arXiv:0806.1964 [hep-th]}}.

\bibitem{Cvetic:2008mh}
M.~Cvetic and T.~Weigand, \emph{{A string theoretic model of gauge mediated
  supersymmetry beaking}},
\href{http://arxiv.org/abs/0807.3953}{{\tt arXiv:0807.3953 [hep-th]}}.

\bibitem{Heckman:2008es}
J.~J. Heckman, J.~Marsano, N.~Saulina, S.~Schafer-Nameki, and C.~Vafa,
  \emph{{Instantons and SUSY breaking in F-theory}},
\href{http://arxiv.org/abs/0808.1286}{{\tt arXiv:0808.1286 [hep-th]}}.

\bibitem{Dudas:2008qf}
E.~Dudas, Y.~Mambrini, S.~Pokorski, A.~Romagnoni, and M.~Trapletti,
  \emph{{Gauge vs. Gravity mediation in models with anomalous U(1)'s}},
  \href{http://dx.doi.org/10.1088/1126-6708/2009/03/011}{JHEP {\bf 03} (2009)
  011},
\href{http://arxiv.org/abs/0809.5064}{{\tt arXiv:0809.5064 [hep-th]}}.

\bibitem{Marsano:2008jq}
J.~Marsano, N.~Saulina, and S.~Schafer-Nameki, \emph{{Gauge Mediation in
  F-Theory GUT Models}},
\href{http://arxiv.org/abs/0808.1571}{{\tt arXiv:0808.1571 [hep-th]}}.

\bibitem{Bianchi:1991eu}
M.~Bianchi, G.~Pradisi, and A.~Sagnotti, \emph{{Toroidal compactification and
  symmetry breaking in open string theories}},
\href{http://dx.doi.org/10.1016/0550-3213(92)90129-Y}{Nucl. Phys. {\bf B376}
  (1992)  365--386}.

\bibitem{Bianchi:1997rf}
M.~Bianchi, \emph{{A note on toroidal compactifications of the type I
  superstring and other superstring vacuum configurations with 16
  supercharges}}, \href{http://dx.doi.org/10.1016/S0550-3213(98)00403-9}{Nucl.
  Phys. {\bf B528} (1998)  73--94},
\href{http://arxiv.org/abs/hep-th/9711201}{{\tt arXiv:hep-th/9711201}}.

\bibitem{Witten:1997bs}
E.~Witten, \emph{{Toroidal compactification without vector structure}}, JHEP
  {\bf 02} (1998)  006,
\href{http://arxiv.org/abs/hep-th/9712028}{{\tt arXiv:hep-th/9712028}}.

\bibitem{Bianchi:2000de}
M.~Bianchi and J.~F. Morales, \emph{{Anomalies and tadpoles}}, JHEP {\bf 03}
  (2000)  030,
\href{http://arxiv.org/abs/hep-th/0002149}{{\tt arXiv:hep-th/0002149}}.

\bibitem{Aldazabal:2000sa}
G.~Aldazabal, L.~E. Ibanez, F.~Quevedo, and A.~M. Uranga, \emph{{D-branes at
  singularities: A bottom-up approach to the string embedding of the standard
  model}}, JHEP {\bf 08} (2000)  002,
\href{http://arxiv.org/abs/hep-th/0005067}{{\tt arXiv:hep-th/0005067}}.

\bibitem{Bianchi:2007fx}
M.~Bianchi and E.~Kiritsis, \emph{{Non-perturbative and Flux superpotentials
  for Type I strings on the orbifold}},
  \href{http://dx.doi.org/10.1016/j.nuclphysb.2007.05.006}{Nucl. Phys. {\bf
  B782} (2007)  26--50},
\href{http://arxiv.org/abs/hep-th/0702015}{{\tt arXiv:hep-th/0702015}}.

\bibitem{Bianchi:2007wy}
M.~Bianchi, F.~Fucito, and J.~F. Morales, \emph{{D-brane Instantons on the
  orientifold}}, \href{http://dx.doi.org/10.1088/1126-6708/2007/07/038}{JHEP
  {\bf 07} (2007)  038},
\href{http://arxiv.org/abs/0704.0784}{{\tt arXiv:0704.0784 [hep-th]}}.

\bibitem{Ibanez:2007tu}
L.~E. Ibanez and A.~M. Uranga, \emph{{Instanton Induced Open String
  Superpotentials and Branes at Singularities}},
  \href{http://dx.doi.org/10.1088/1126-6708/2008/02/103}{JHEP {\bf 02} (2008)
  103},
\href{http://arxiv.org/abs/0711.1316}{{\tt arXiv:0711.1316 [hep-th]}}.

\bibitem{Shifman:1995ua}
M.~A. Shifman, \emph{{Nonperturbative dynamics in supersymmetric gauge
  theories}}, \href{http://dx.doi.org/10.1016/S0146-6410(97)00042-2}{Prog.
  Part. Nucl. Phys. {\bf 39} (1997)  1--116},
\href{http://arxiv.org/abs/hep-th/9704114}{{\tt arXiv:hep-th/9704114}}.

\bibitem{Shifman:1999mv}
M.~A. Shifman and A.~I. Vainshtein, \emph{{Instantons versus supersymmetry:
  Fifteen years later}},
\href{http://arxiv.org/abs/hep-th/9902018}{{\tt arXiv:hep-th/9902018}}.

\bibitem{Peskin:1997qi}
M.~E. Peskin, \emph{{Duality in supersymmetric Yang-Mills theory}},
\href{http://arxiv.org/abs/hep-th/9702094}{{\tt arXiv:hep-th/9702094}}.

\bibitem{Terning:2006bq}
J.~Terning, \emph{{Modern supersymmetry: Dynamics and duality}},. Oxford, UK:
  Clarendon (2006) 324 p.

\bibitem{Angelantonj:1996uy}
C.~Angelantonj, M.~Bianchi, G.~Pradisi, A.~Sagnotti, and Y.~S. Stanev,
  \emph{{Chiral asymmetry in four-dimensional open- string vacua}},
  \href{http://dx.doi.org/10.1016/0370-2693(96)00869-6}{Phys. Lett. {\bf B385}
  (1996)  96--102},
\href{http://arxiv.org/abs/hep-th/9606169}{{\tt arXiv:hep-th/9606169}}.

\bibitem{Anastasopoulos:2006cz}
P.~Anastasopoulos, M.~Bianchi, E.~Dudas, and E.~Kiritsis, \emph{{Anomalies,
  anomalous U(1)'s and generalized Chern-Simons terms}}, JHEP {\bf 11} (2006)
  057,
\href{http://arxiv.org/abs/hep-th/0605225}{{\tt arXiv:hep-th/0605225}}.

\bibitem{Akerblom:2006hx}
N.~Akerblom, R.~Blumenhagen, D.~Lust, E.~Plauschinn, and M.~Schmidt-Sommerfeld,
  \emph{{Non-perturbative SQCD Superpotentials from String Instantons}}, JHEP
  {\bf 04} (2007)  076,
\href{http://arxiv.org/abs/hep-th/0612132}{{\tt arXiv:hep-th/0612132}}.

\bibitem{Argurio:2007vqa}
R.~Argurio, M.~Bertolini, G.~Ferretti, A.~Lerda, and C.~Petersson,
  \emph{{Stringy Instantons at Orbifold Singularities}}, JHEP {\bf 06} (2007)
  067,
\href{http://arxiv.org/abs/0704.0262}{{\tt arXiv:0704.0262 [hep-th]}}.

\bibitem{Blumenhagen:2006xt}
R.~Blumenhagen, M.~Cvetic, and T.~Weigand, \emph{{Spacetime instanton
  corrections in 4D string vacua - the seesaw mechanism for D-brane models}},
  \href{http://dx.doi.org/10.1016/j.nuclphysb.2007.02.016}{Nucl. Phys. {\bf
  B771} (2007)  113--142},
\href{http://arxiv.org/abs/hep-th/0609191}{{\tt arXiv:hep-th/0609191}}.

\bibitem{Ibanez:2006da}
L.~E. Ibanez and A.~M. Uranga, \emph{{Neutrino Majorana masses from string
  theory instanton effects}}, JHEP {\bf 03} (2007)  052,
\href{http://arxiv.org/abs/hep-th/0609213}{{\tt arXiv:hep-th/0609213}}.

\bibitem{Ibanez:2007rs}
L.~E. Ibanez, A.~N. Schellekens, and A.~M. Uranga, \emph{{Instanton Induced
  Neutrino Majorana Masses in CFT Orientifolds with MSSM-like spectra}}, JHEP
  {\bf 06} (2007)  011,
\href{http://arxiv.org/abs/0704.1079}{{\tt arXiv:0704.1079 [hep-th]}}.

\bibitem{Billo':2008sp}
M.~Billo' {\em et al.}, \emph{{Flux interactions on D-branes and instantons}},
  \href{http://dx.doi.org/10.1088/1126-6708/2008/10/112}{JHEP {\bf 10} (2008)
  112},
\href{http://arxiv.org/abs/0807.1666}{{\tt arXiv:0807.1666 [hep-th]}}.

\bibitem{Billo':2008pg}
M.~Billo' {\em et al.}, \emph{{Non-perturbative effective interactions from
  fluxes}}, \href{http://dx.doi.org/10.1088/1126-6708/2008/12/102}{JHEP {\bf
  12} (2008)  102},
\href{http://arxiv.org/abs/0807.4098}{{\tt arXiv:0807.4098 [hep-th]}}.

\end{thebibliography}
\providecommand{\href}[2]{#2}\begingroup\raggedright\endgroup

\end{document}